\documentclass[%
 reprint,
superscriptaddress,
nofootinbib,
 amsmath,amssymb,
 aps,
prl,
]{revtex4-2}
\usepackage[utf8]{inputenc}
\usepackage{amsmath}
\usepackage{amsfonts}
\usepackage{amssymb}
\usepackage{graphicx}
\usepackage{hyperref}
\usepackage{braket}
\usepackage[colorinlistoftodos]{todonotes}
\usepackage[left=2cm,right=2cm,top=2cm,bottom=2cm]{geometry}
\usepackage[caption=false]{subfig}
\usepackage{float}
\usepackage[T1]{fontenc}

\begin{document}

\title{Inferring charge noise source locations from correlations in spin qubits}

\author{J. S. Rojas-Arias}
\email{juan.rojasarias@riken.jp}
\affiliation{RIKEN, Center for Quantum Computing (RQC), Wako-shi, Saitama 351-0198, Japan}
\author{A. Noiri}
\affiliation{RIKEN, Center for Emergent Matter Science (CEMS), Wako-shi, Saitama 351-0198, Japan}
\author{J. Yoneda}
\affiliation{Department of Advanced Materials Science, University of Tokyo, Kashiwa, Chiba 277-8561, Japan}
\author{P. Stano}
\affiliation{RIKEN, Center for Quantum Computing (RQC), Wako-shi, Saitama 351-0198, Japan}
\affiliation{Slovak Academy of Sciences, Institute of Physics, 845 11 Bratislava, Slovakia}
\author{T. Nakajima}
\affiliation{RIKEN, Center for Emergent Matter Science (CEMS), Wako-shi, Saitama 351-0198, Japan}
\author{K. Takeda}
\affiliation{RIKEN, Center for Emergent Matter Science (CEMS), Wako-shi, Saitama 351-0198, Japan}
\author{T. Kobayashi}
\affiliation{RIKEN, Center for Quantum Computing (RQC), Wako-shi, Saitama 351-0198, Japan}
\author{G. Scappucci}
\affiliation{QuTech and Kavli Institute of Nanoscience, Delft University of Technology, Lorentzweg 1, 2628 CJ Delft, Netherlands}

\author{S. Tarucha}
\email{tarucha@riken.jp}
\affiliation{RIKEN, Center for Emergent Matter Science (CEMS), Wako-shi, Saitama 351-0198, Japan}
\affiliation{RIKEN, Center for Quantum Computing (RQC), Wako-shi, Saitama 351-0198, Japan}%
\author{D. Loss}
\affiliation{RIKEN, Center for Quantum Computing (RQC), Wako-shi, Saitama 351-0198, Japan}
\affiliation{Department of Physics, University of Basel, Klingelbergstrasse 82, CH-4056 Basel, Switzerland}

\begin{abstract}

We investigate low-frequency noise in a spin-qubit device made in isotopically purified Si/Si-Ge. Observing sizable cross-correlations among energy fluctuations of different qubits, we conclude that these fluctuations are dominated by charge noise. At low frequencies, the noise spectra are not well described by a power law; instead, they reveal the presence of a few individual two-level fluctuators (TLFs). We demonstrate that the noise cross-correlations allow one to get information on the spatial location of such individual TLFs.

\end{abstract}

\maketitle

\section{Introduction}

Charge noise is one of the major factors limiting the performance of solid-state quantum processors. The $1/f$-like power spectrum often observed at low frequencies is believed to arise from an ensemble of two-level 
fluctuators (TLFs) \cite{Bernamont1937}. In silicon spin qubits, particularly those where the hyperfine noise is reduced by isotopic purification, the individual TLFs can sometimes be identified \cite{Rojas-Arias2023,Connors2022}. Recent work \cite{Ye2024} has further shown that temperature and gate voltages can strongly affect the switching times of individual TLFs.

Despite extensive research, the precise nature of TLFs remains elusive \cite{Muller2019}. Various suggestions predict TLFs located inside oxides, at interfaces \cite{Connors2019,Kuhlmann2013}, or at the quantum well \cite{PaqueletWuetz2023}. In bare wafers, one can look for TLFs and impurities by direct imaging methods, such as atomic force microscopy; for example, the in-plane locations of TLFs at the Si/SiO$_2$ interface were traced in Ref.~\cite{Cowie2024}, and the 2DEG potential was mapped in Ref.~\cite{Topinka2001}. However, for an actual nanodevice with a complex multilayer gate structure, direct imaging is difficult, and any hints on the spatial locations of TLFs causing qubit decoherence would be valuable.

In our recent works, we have developed \cite{Gutierrez-Rubio2022} and applied \cite{Yoneda2023,Rojas-Arias2023} cross-power spectral densities (cross-PSDs) as a tool for the experimental study of noise in spin qubits \cite{Donnelly2025}. The long-distance tails of noise correlations play a key role in quantum error correction \cite{Clader2021,Clemens2004}. At the same time, cross-correlations reveal temporal and spatial noise features that are absent in auto-correlations In this Letter, we demonstrate that cross-PSDs can be used to estimate the spatial location of a TLF, on top of its characteristic switching time visible in the auto-PSDs.

\section{The essence of the triangulation method}

\begin{figure}
\centering
\includegraphics[width=0.65\columnwidth]{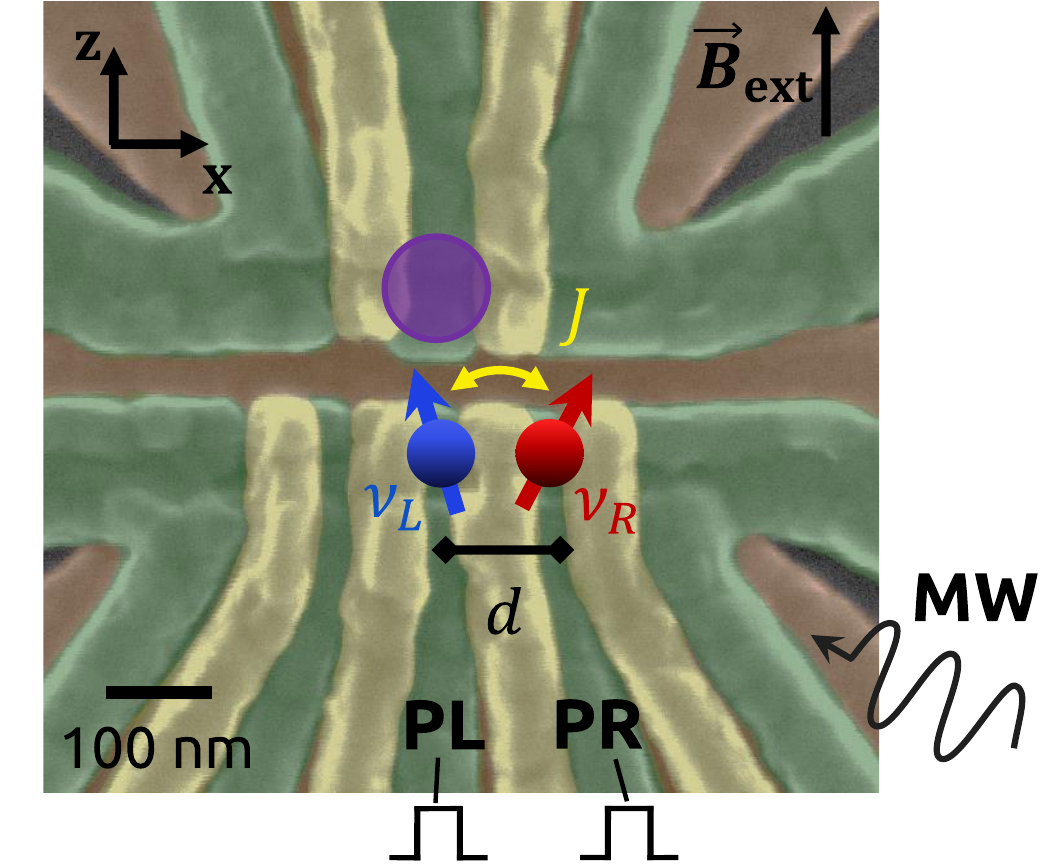}
\caption{Image of a device identical to the one measured. Qubits, depicted as colored spheres with arrows, are formed under plunger gates PL and PR. The qubits are separated by distance $d$, have Zeeman energies $h\nu_L$ and $h\nu_R$, and interact via exchange $hJ$. The  purple circle indicates the charge sensor.
}
\label{fig:device}
\end{figure}

\begin{figure*}[tbp]
\centering
\subfloat{\includegraphics[width=0.95\textwidth]{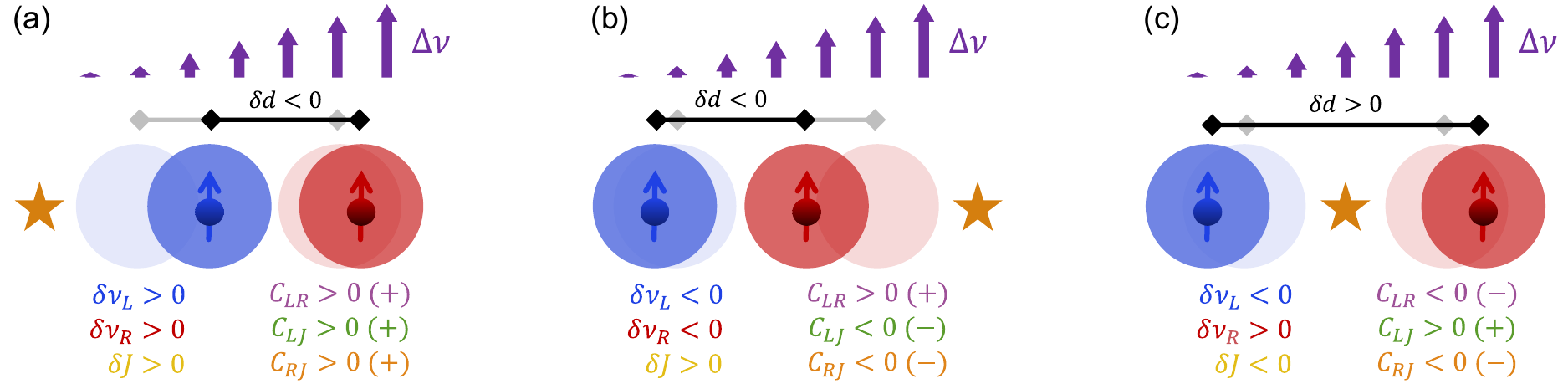}\label{fig:loc_left}}
\subfloat{\label{fig:loc_right}}
\subfloat{\label{fig:loc_middle}}
\caption{Schematics of a TLF located (a) left, (b) right, or (c) between two qubits. The left (right) qubit is shown as a blue (red) arrow in a circle, and the TLF as a dark orange star. A purple gradient marks increasing qubit energies to the right, with the black line indicating qubit separation. Faint colors show positions before the $|0\rangle \to |1\rangle$ TLF switch. Bottom panels display the resulting energy shifts and cross-PSD signs due to the TLF.}
\label{fig:location}
\end{figure*}

To explain the essence of the method, we start with the one-dimensional model of Ref.~\cite{Yoneda2023}, describing two spin qubits coupled via exchange $J$ in the device of Fig.~\ref{fig:device}. The external field $\mathbf{B}_\mathrm{ext}$ lies in the 2DEG plane along $z$, and a micromagnet provides the largest $B_z$ gradient along the array axis $x$.

Charge noise shifts the qubits within the plane. In the inhomogeneous field, these shifts induce Zeeman-energy noise,
\begin{align}
\delta\nu_\alpha=\frac{g\mu_B}{h}\frac{\partial B_z}{\partial x}\delta x_\alpha,
\label{eq:fluc_freq}
\end{align}
for qubit $\alpha\in{L,R}$, with $\delta x_\alpha$ the displacement along $x$, $h$ Planck's constant, $\mu_B$ Bohr's magneton, and $g$ the electron $g$-factor. We assume (i) $\partial B_z/\partial x$ is positive and equal for both qubits, (ii) other magnetic gradients are negligible, (iii) the two dots in-plane confinements are similar, and (iv) the dots operate at the charge symmetry point \cite{Martins2016,Reed2016}. These assumptions are motivated by the device or serve only to simplify the presentation.

At the symmetry point, exchange fluctuations depend only on inter-dot tunneling, and hence  a function of the inter-dot separation $d=x_R-x_L$. To lowest order, we have
\begin{align}
\delta J=\frac{\partial J}{\partial d}(\delta x_R-\delta x_L).
\label{eq:fluc_exch}
\end{align}

Equations~\eqref{eq:fluc_freq} and \eqref{eq:fluc_exch} allow to relate exchange PSDs to qubit-energy PSDs, via
\begin{subequations}
\begin{align}
S_J(f)=\left(\frac{\partial J}{\partial \Delta}\right)^2[S_L(f)+S_R(f)-2\mathfrak{Re}C_{LR}(f)],\\
C_{LJ}(f)=-\frac{\partial J}{\partial\Delta}[S_L(f)-C_{LR}(f)],\\
C_{RJ}(f)=\frac{\partial J}{\partial\Delta}[S_R(f)-C_{LR}^*(f)].
\end{align}
\label{eq:corr_exch}%
\end{subequations}
with $\Delta=\nu_R-\nu_L$. We have denoted an auto-PSD by $S_\beta(f)$ and a cross-PSD by $C_{\beta \beta^\prime}(f)$, with shorthands $\nu_\beta \to \beta$ for the PSD indexes of quantities $\beta \in \{\nu_L,\nu_R,J\}$. Thus, out of the six a-priori independent PSDs for $\{\nu_L,\nu_R,J\}$, three are eliminated by Eq.~\eqref{eq:corr_exch}.

To explain how a noise source location can be inferred from cross-PSDs, we look at three scenarios drawn in Fig.~\ref{fig:location}, with a TLF to the left, in between, or to the right of the qubits. While a TLF could in principle lie anywhere perpendicular to the qubit axis, here we focus on its position along the axis, assuming the magnetic-field gradient is dominant there and neglecting perpendicular contributions. We label the TLF states so that $\ket{1}$ electrostatically repels the qubits more than $\ket{0}$. We analyze the $\ket{0}\to \ket{1}$ transition, noting that the cross-PSD signs are unchanged for the reverse process and thus independent of labeling. We also assume the TLF acts instantaneously on both qubits, making all cross-PSDs real, as observed experimentally \cite{Yoneda2023,Rojas-Arias2023,Rojas-Arias2024}.

First, let us take a TLF located left of the array, Fig.~\ref{fig:loc_left}. The TLF switch $\ket{0}\rightarrow\ket{1}$ pushes qubits to the right and, with a positive magnetic gradient, both qubit energies increase. The left qubit shifts slightly more, since it is closer to the TLF. Hence, the inter-dot separation decreases and the exchange goes up. In summary, all three quantities increase, and all three cross-PSDs are positive.

Second, consider a TLF located to the right of the array, Fig.~\ref{fig:loc_right}. The TLF switch pushes the qubits in the opposite direction, and their energies decrease. The inter-dot distance still drops, in the same way as in the previous scenario. In summary, the two qubit energies are positively correlated, while each of them is negatively correlated with exchange.

Finally, consider a TLF inside the array, in between the two dots. The TLF switch pushes the dots apart, decreasing the left qubit energy while increasing the right qubit energy. Also, only in this case the inter-dot distance increases and the exchange energy drops. These changes result in the cross-correlations as given in Fig.~\ref{fig:loc_middle}.

A quantum device is typically influenced by many TLFs with different switching times and random locations, whose additive effects are hard to disentangle. Still, individual TLFs can sometimes be identified in the spectra as single Lorentzian contributions, from which their switching times are extracted. The procedure above goes further by also revealing their spatial location, providing information inaccessible from auto-PSDs alone.

\section{Experimental demonstration}\label{sec:device}

We demonstrate the method on data from the device in Fig.~\ref{fig:device}, previously used in Refs.~\cite{Noiri2022,Rojas-Arias2023}. It implements a triple quantum dot with aluminum gates over isotopically purified Si/Si-Ge (800 ppm $^{29}$Si), where we operate only the two dots under PL and PR as single-electron spin qubits. We obtain time traces of $\nu_L$, $\nu_R$, and $J$  via repeated interleaved Ramsey experiments \cite{Yoneda2023}. In the Supplemental Material (SM) \cite{SM}, we provide details on the experiment and data analysis.

\begin{figure}[htbp]
\centering
\includegraphics[width=0.8\columnwidth]{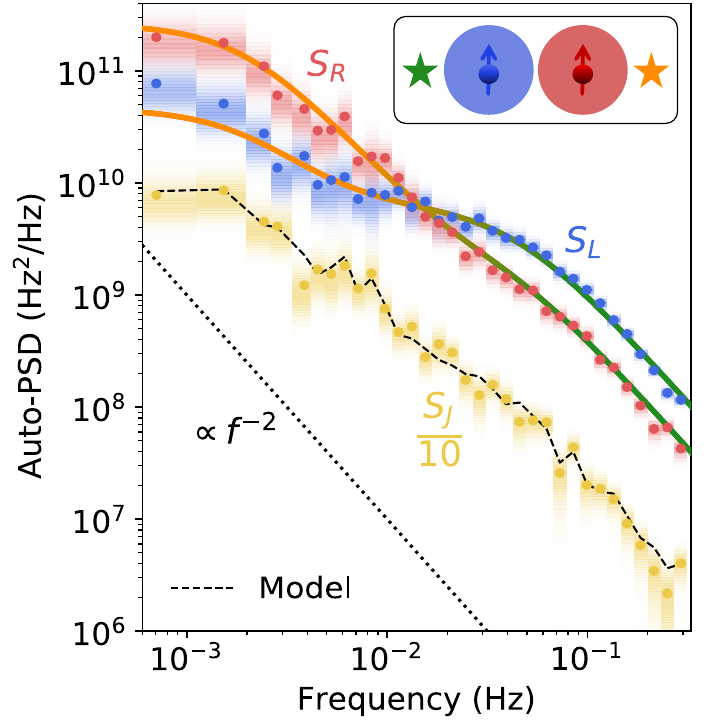}
\caption{Auto-PSDs of $S_L$ (blue), $S_R$ (red), and $S_J$ (yellow). Color gradients show Bayesian probability distributions \cite{Gutierrez-Rubio2022}, with means as points. $S_J$ is shifted down by a factor 10 for clarity. The black dashed line is the model prediction from Eq.~\eqref{eq:corr_exch}. Gradient-colored fits of $S_L(f)$ and $S_R(f)$ use two Lorentzians $\propto [1+4\pi^2f^2\tau^2]^{-1}$ (see SM), with switching times $\tau=3.7$~s (green) and $\tau=90$~s (orange). A dotted line shows an $f^{-2}$ dependence for reference.
}
\label{fig:auto-psds}
\end{figure}

Looking at the auto-PSDs in Fig.~\ref{fig:auto-psds}, two facts stand out: $S_L(f)$ and $S_R(f)$ have different shapes, and neither follows a simple $1/f$ law. These observations suggest that the noise arises from a few local sources near the qubits rather than from a global property. Indeed, the features can be captured by only two dominant TLFs, whose contributions appear as individual Lorentzians (solid lines), consistent with previous reports \cite{Rojas-Arias2023, Connors2022, Ye2024}. As we show next, the same two TLFs also reproduce the structure of the cross-PSDs, providing a unified picture of correlated noise. The SM presents the full set of fits, including all auto- and cross-PSDs.

\begin{figure*}
\centering
\subfloat{\includegraphics[width=0.95\textwidth]{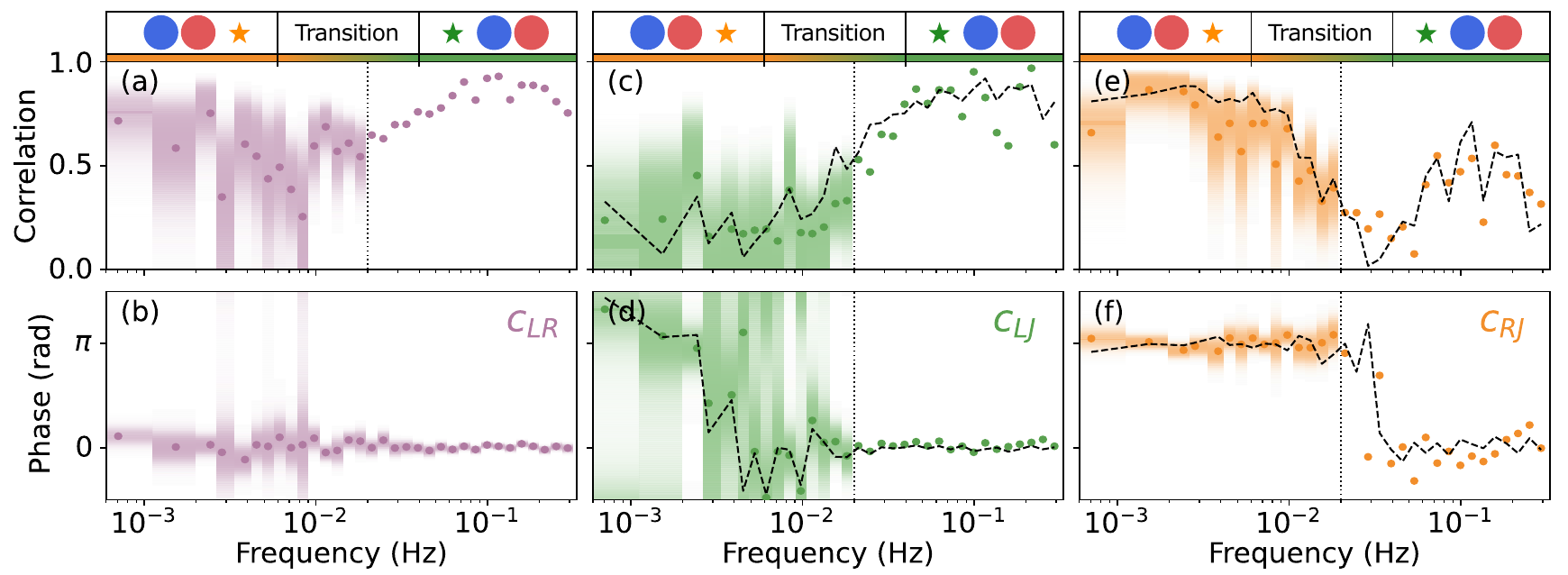}\label{fig:clr_s}}
\subfloat{\label{fig:clr_p}}
\subfloat{\label{fig:clj_s}}
\subfloat{\label{fig:clj_p}}
\subfloat{\label{fig:crj_s}}
\subfloat{\label{fig:crj_p}}
\caption{Normalized cross-PSDs for the two-qubit system: $c_{LR}$ (purple), $c_{LJ}$ (green), and $c_{RJ}$ (orange). Magnitudes are shown in (a,c,e) and phases in (b,d,f). Color gradients give Bayesian probability distributions with means as points. Vertical dotted lines mark the frequency above which spectra are corrected for estimation errors (see SM); only the phase of $c_{LR}$ requires no correction. Black dashed lines are predictions from Eq.~\eqref{eq:corr_exch}. Schematics indicate the dominant TLF regime: right of the qubits at low frequencies, left at high, and a transition in between.}
\label{fig:cross-psds}
\end{figure*}

To quantify correlations, we use cross-PSDs normalized by the auto-PSDs, $c_{\alpha\beta}(f)=C_{\alpha\beta}(f)/\sqrt{S_\alpha(f)S_\beta(f)}$. Each is a complex number with magnitude between 0 (uncorrelated) and 1 (perfectly correlated) and a phase. Figures~\ref{fig:clr_s} and \ref{fig:clr_p} show the magnitudes and phases for qubit-qubit correlations. The correlations exceed $50\%$ across most frequencies and reach over $90\%$ at some points. The phase is zero everywhere, allowing us to exclude the scenario of Fig.~\ref{fig:loc_middle}. Thus, the two dominant TLFs in Fig.~\ref{fig:auto-psds} are located outside the dot array.

We next examine correlations between qubit energies and exchange, shown in Figs.~\ref{fig:clj_s}--\ref{fig:crj_p}. The magnitudes of $c_{LJ}(f)$ and $c_{RJ}(f)$ often exceed $80\%$. Since exchange arises purely from wavefunction overlap, such strong correlations demonstrate that the qubit-energy fluctuations have an electrical origin. This agrees with earlier results in Si/Si-Ge qubits \cite{Connors2022, Rojas-Arias2023, Yoneda2023, Ye2024}, and contrasts with Si/SiO$_2$ devices where the relevant TLFs are individual nuclear spins \cite{Zhao2019,Hensen2020,Huang2019}. We further test consistency with Eq.~\eqref{eq:corr_exch}, with the dashed lines in Figs.~\ref{fig:auto-psds} and \ref{fig:cross-psds} showing predictions obtained without free parameters, using $\partial J/\partial\Delta=-0.87$ extracted independently from $J$ and $\Delta$ fluctuations (see SM). The excellent agreement confirms our noise model (Eq.~\eqref{eq:fluc_freq}) and shows that individual TLFs can be resolved in both time traces and PSDs, enabling the classification of Fig.~\ref{fig:location}.

The phases of the three cross-PSDs, shown in the bottom row of Fig.~\ref{fig:cross-psds}, reveal three regions. For $f<6$ mHz, they match Fig.~\ref{fig:loc_right}, indicating a TLF to the right of the array, consistent with the stronger noise of the right qubit in Fig.~\ref{fig:auto-psds}. For $f>40$~mHz, they correspond to Fig.~\ref{fig:loc_left}, with a TLF to the left, again corroborated by stronger noise in the left qubit. A two-TLF fit (see SM) assigns switching times of 90~s and 3.7~s to these fluctuators, respectively. The intermediate region, $6 \leq f \leq 40$~mHz, reflects comparable effects of both TLFs and is a transition regime; the observed phases do not correspond to any scenario in Fig.~\ref{fig:location}.

\section{Discussion}

We now discuss our results focusing on applicability and limitations. In particular, we (i) apply our method to an independent data set, (ii) consider extensions to larger arrays, (iii) examine refined triangulation through microscopic TLF models, (iv) outline potential problems, and (v) discuss other spin-qubit encodings.

{First, to assess its generality, we apply our method to data obtained in Ref.~\cite{Yoneda2023} (natural silicon device with a different gate layout). The analysis (shown in SM) reveals two TLFs, one of which is positioned between the qubits (Fig.~\ref{fig:loc_middle}), the other to the left. Including these data, we thus observe all three scenarios  of Fig.~\ref{fig:location} across four detectable TLFs, suggesting that each scenario occurs frequently.}

\begin{figure}
\centering
\subfloat{\includegraphics[width=0.90\columnwidth]{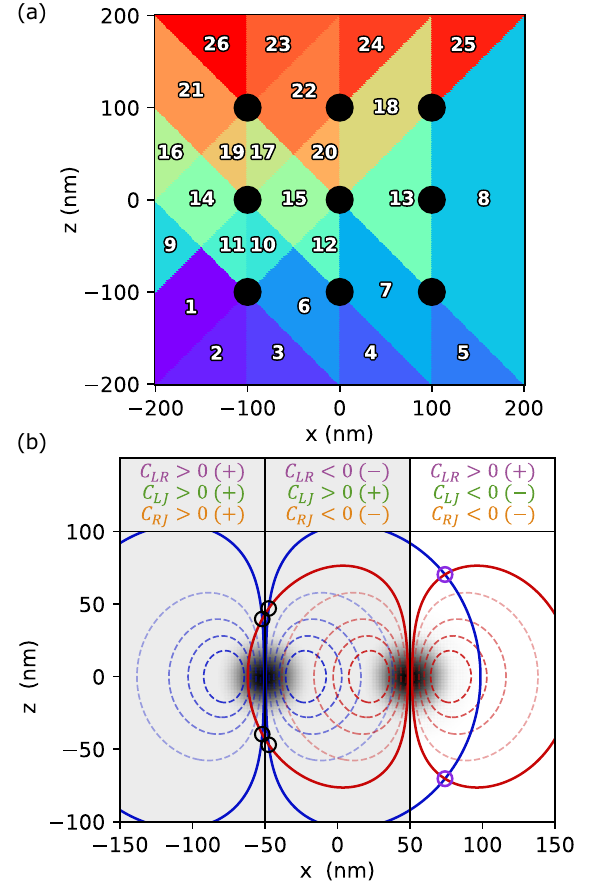}\label{fig:tangram}}
\subfloat{\label{fig:rings}}
\caption{TLF triangulation in two dimensions. (a) Regions with distinct cross-PSD signatures for a $3\times3$ qubit array (black circles). The tiling corresponds to a magnetic gradient $\nabla B_z \propto 1\hat{x} - 0.1 z \hat{z}$, similar to that of the device in Fig.~\ref{fig:device} \cite{Rojas-Arias2023}. Analogous tilings can be constructed for any known magnetic profile (or spin-orbit axis or $g$-tensor). (b) Contours of qubit-TLF interaction strength for a screened charge trap \cite{Rojas-Arias2023}. Qubit wavefunctions are shown as Gaussian profiles (20 nm width). Dashed blue (red) lines denote contours of constant energy shifts for the left (right) qubit; solid lines match fitted jumps from PSD analysis ($|g^L|=14.9$ kHz, $|g^R|=38.0$ kHz, for the 90~s TLF). Intersections mark six candidate positions, with cross-PSD signs excluding the grey region and only leaving two possible locations (purple circles).}
\label{fig:extension}
\end{figure}

{Second, our analysis naturally extends to larger qubit arrays \cite{Philips2022}, as we show in simulations. For two qubits, correlations with exchange were needed to distinguish whether a TLF lies to the left or right of the qubits. In larger arrays, qubit-qubit energy correlations alone suffice, eliminating the need for exchange correlations and improving scalability. An illustrative case is shown in Fig.~\ref{fig:tangram} for a $3\times3$ lattice. Knowing the magnetic-field profile, the analysis of Fig.~\ref{fig:location} generalizes to two dimensions (see SM), where the plane divides into regions with distinct cross-PSD signatures, yielding a “tangram”-like tiling (Fig.~\ref{fig:tangram}). Once a clear TLF signature is identified, its position is restricted to one such regions.}

{Third, we turn to refined triangulation based on microscopic modeling. At this stage it is useful to separate two levels of localization: from the signs of the cross-PSDs we obtain a qualitative classification that places a TLF relative to the array (Figs.~\ref{fig:location} and \ref{fig:tangram}), whereas extracting quantitative positions or absolute distances requires a model for the microscopic nature of the defect. If the spatial scaling of the TLF-qubit interaction were known, finding the qubit-TLF coupling $g^\alpha$ for several qubits could, in principle, reveal the TLF's position. Even though measurement uncertainties usually limit such pinpointing, for the sake of illustration, we revisit the approach of Ref.~\cite{Rojas-Arias2023} and model TLFs as charge traps in the oxide above the 2DEG, screened by a nearby metal. The $\ket{0}\to\ket{1}$ transition corresponds to trapping an elementary charge, equivalent to a dipole oriented perpendicular to the 2DEG. This model narrows the possible locations, illustrated in Fig.~\ref{fig:rings} for the 90~s TLF (see SM). Namely, the qubit-TLF couplings obtained from the PSD fits locate the TLF to the intersection of the solid curves (contours of constant qubit-energy shifts), yielding six candidate positions. The qualitative classification from cross-PSD signs (Fig.~\ref{fig:location}) excludes four and thus reveals that the TLF is horizontally about 100 nm from the qubit-array center. Nevertheless, we stress that extracted positions are sensitive to modeling assumptions. They depend on whether the TLF is an elementary charge or dipole, its distance to the 2DEG and screening metal, and the dipole’s size and orientation. Without detailed knowledge or well-justified assumptions, absolute localization thus remains elusive. On the other hand, the same sensitivity helps narrow down the TLF model, as it should consistently describe spectra measured for multiple qubits simultaneously.
}

{Fourth, we address possible ambiguities in interpreting cross-PSDs. In the previous paragraph, we considered a dipolar TLF with its charge dipole pointing out of the 2DEG plane. Were it oriented along the array axis, its effects would be more complex. For example, if located between the qubits, a TLF-state switch could attract one qubit while repelling the other, producing signatures that mimic a TLF pointing out of the plane located on either side of the array. Nevertheless, this ambiguity could be resolved by examining neighboring qubits, since their coupling strengths would differ strongly depending on the true TLF location. 
In contrast, a negative qubit-qubit cross-PSD, $C_{LR}(f)<0$ (observed in the SM), can only arise from a TLF between the qubits and is therefore a robust signature of such scenario. A similar ambiguity arises when the dots’ in-plane confinements strongly differ. Then, a TLF switch could displace the loosely bound dot more, even if it is farther from the TLF. However, such asymmetry would yield smaller noise in the tightly bound dot, inconsistent with the crossover between $S_L$ and $S_R$ in Fig.~\ref{fig:auto-psds}. Finally, we consider the case of two (or more) TLFs having similar switching times. Their Lorentzian spectra overlap and, in two-qubit data, cannot be distinguished from a single TLF. Data from larger arrays, however, could help identify individual TLFs by their distinct spatial correlation patterns. 
We conclude with a somewhat obvious statement: the more precisely the dots and TLFs are characterized, the more reliable the classification of Fig.~\ref{fig:location} and associated PSD fits will be.}

{Finally, we comment on the broader applicability of our approach to other spin-qubit encodings. Although we focused on electron spins in micromagnet-induced gradients, the same principles apply to other spin-qubit encodings. For hole spin qubits in germanium, anisotropic $g$-tensors replace engineered gradients; once the $g$-tensor axes are known, the tangram-like classification of Fig.~\ref{fig:tangram} can be constructed and TLFs localized from cross-PSD signs. A similar strategy applies to singlet-triplet or exchange-only (EO) qubits, using cross-PSDs of exchange interactions. For example, in a linear EO qubit formed by three spins with intra-qubit couplings $J_{12}$ and $J_{23}$, a TLF on either side of the array yields positive correlations between the two exchanges, while one inside gives negative correlations. Inter-qubit couplings to neighboring qubits, or triangular EO geometries with additional intra-qubit coupling $J_{13}$ \cite{Acuna2024}, could provide extra constraints to further refine localization. Either way, EO qubit lattices map onto tangram-like patterns of exchange cross-PSD signs (see SM), providing a general framework to infer TLF positions.}

\section{Conclusions}\label{sec:conc}

Using two spin qubits in an isotopically purified Si/Si-Ge device, we have investigated noise cross-correlations between the qubit Zeeman energies and between the qubit energies and exchange energy, finding them consistent with charge TLFs located near the qubits. We introduced a simple classification where the signs of the three cross-PSDs reveal the spatial location of dominant TLFs and their characteristic times. In our device, we identified one TLF to the left of the array with a switching time of 3.7~s and another to the right with a switching time of 90~s. Extending the method to two dimensions, we constructed tiling plots that divide real space into regions with distinct cross-PSD signatures. This framework provides a systematic way to classify possible TLF locations in larger qubit arrays and generalizes naturally to other spin-qubit encodings.

\acknowledgments
This work was supported by the Swiss National Science Foundation and NCCR SPIN Grant No. 51NF40-180604, JST Moonshot R\&D Grant Numbers JPMJMS226B and JPMJMS2065, JSPS KAKENHI Grant Nos. JP23H01790 and JP23H05455. J.S.R.-A. acknowledges support from the Gutaiteki Collaboration Seed, and J.Y. support from The Precise Measurement Technology Promotion Foundation and The Nakajima Foundation. We also acknowledge support from JST PRESTO Grant Numbers JPMJPR23F8 (A.N.), JPMJPR2017 (T.N.) and JPMJPR21BA (J.Y.).

\bibliography{references.bib}

\widetext
\clearpage

\begin{center}
\textbf{\large Supplemental Material for ``Inferring the location of charge noise sources acting on spin qubits''}
\end{center}

\setcounter{equation}{0}
\setcounter{figure}{0}
\setcounter{table}{0}
\setcounter{page}{1}

\renewcommand{\theequation}{S\arabic{equation}}
\renewcommand{\thefigure}{S\arabic{figure}}

\title{Supplemental Material for ``Inferring the location of charge noise sources acting on spin qubits''}

\author{J. S. Rojas-Arias}
\email{juan.rojasarias@riken.jp}
\affiliation{RIKEN, Center for Quantum Computing (RQC), Wako-shi, Saitama 351-0198, Japan}
\author{A. Noiri}
\affiliation{RIKEN, Center for Emergent Matter Science (CEMS), Wako-shi, Saitama 351-0198, Japan}
\author{J. Yoneda}
\affiliation{Department of Advanced Materials Science, University of Tokyo, Kashiwa, Chiba 277-8561, Japan}
\author{P. Stano}
\affiliation{RIKEN, Center for Quantum Computing (RQC), Wako-shi, Saitama 351-0198, Japan}
\affiliation{Slovak Academy of Sciences, Institute of Physics, 845 11 Bratislava, Slovakia}
\author{T. Nakajima}
\affiliation{RIKEN, Center for Emergent Matter Science (CEMS), Wako-shi, Saitama 351-0198, Japan}
\author{K. Takeda}
\affiliation{RIKEN, Center for Emergent Matter Science (CEMS), Wako-shi, Saitama 351-0198, Japan}
\author{T. Kobayashi}
\affiliation{RIKEN, Center for Quantum Computing (RQC), Wako-shi, Saitama 351-0198, Japan}
\author{G. Scappucci}
\affiliation{QuTech and Kavli Institute of Nanoscience, Delft University of Technology, Lorentzweg 1, 2628 CJ Delft, Netherlands}

\author{S. Tarucha}
\email{tarucha@riken.jp}
\affiliation{RIKEN, Center for Emergent Matter Science (CEMS), Wako-shi, Saitama 351-0198, Japan}
\affiliation{RIKEN, Center for Quantum Computing (RQC), Wako-shi, Saitama 351-0198, Japan}%
\author{D. Loss}
\affiliation{RIKEN, Center for Quantum Computing (RQC), Wako-shi, Saitama 351-0198, Japan}
\affiliation{Department of Physics, University of Basel, Klingelbergstrasse 82, CH-4056 Basel, Switzerland}


\maketitle


\setcounter{equation}{0}
\setcounter{figure}{0}
\setcounter{table}{0}
\setcounter{page}{1}

\renewcommand{\theequation}{S\arabic{equation}}
\renewcommand{\thefigure}{S\arabic{figure}}

\section{Definition of Power Spectral Densities}

For a pair of fluctuating signals $\delta x(t)$ and $\delta y(t)$, we work with the power spectral density (PSD) defined as:
\begin{align}
C_{\alpha\beta}(f)=\int_{-\infty}^{\infty}dt\ e^{2\pi i f t}\braket{\delta\alpha(0)\delta\beta(t)},
\end{align}
with $\alpha,\beta\in\{x,y\}$. When $\alpha\neq\beta$, $C_{\alpha\beta}(f)$ corresponds to the cross-PSD. The auto-PSDs correspond to the case of individual signals $S_\alpha(f)\equiv C_{\alpha\alpha}(f)$.

\section{Experimental details}\label{app:exp}

The device used in this Letter, shown in Fig.~1 in the main text, consists of a triple quantum dot design made of aluminum gates over an isotopically purified Si/Si-Ge heterostructure (800 ppm $^{29}$Si isotopes). Throughout the experiment we only form quantum dots under plunger gates PL and PR. We work in the single-electron regime in each quantum dot and define qubits in the spin state of the electrons by applying an in-plane external magnetic field $B_\mathrm{ext}=403$ mT. Charge sensing is done via radiofrequency reflectometry coupling a sensor quantum dot (purple circle in Fig.~1) near the qubits to a tank circuit \cite{Noiri2020}. Energy-selective tunneling is used to readout the spin state in a single-shot manner from the charge information \cite{Elzerman2004}. A micromagnet on top provides an artificial spin-orbit interaction that allows for electrical addressing of the spins \cite{Takeda2016}. The device is cooled in a dilution refrigerator operating at 10 mK.

The interacting two-qubit system is determined by three quantities: the Zeeman energies of the left (right) qubit $h\nu_L$ ($h\nu_R$), and exchange interaction $hJ$, with $h$ Planck's constant. Due to the exchange coupling, each qubit has two associated energies depending on whether the other qubit is in the spin-up or spin-down state, leading to a total of four energies that are a function of the set $\{\nu_L,\nu_R,J\}$. We label $h\nu_\alpha^\sigma$ the energy of qubit $\alpha\in\{L,R\}$ when the other qubit is in the $\sigma\in\{\uparrow,\downarrow\}$ state. To obtain estimates of these quantities we perform a repeated interleaved Ramsey experiment as was done in Ref.~\cite{Yoneda2023}.

\begin{figure}[htbp]
\centering
\includegraphics[width=0.45\columnwidth]{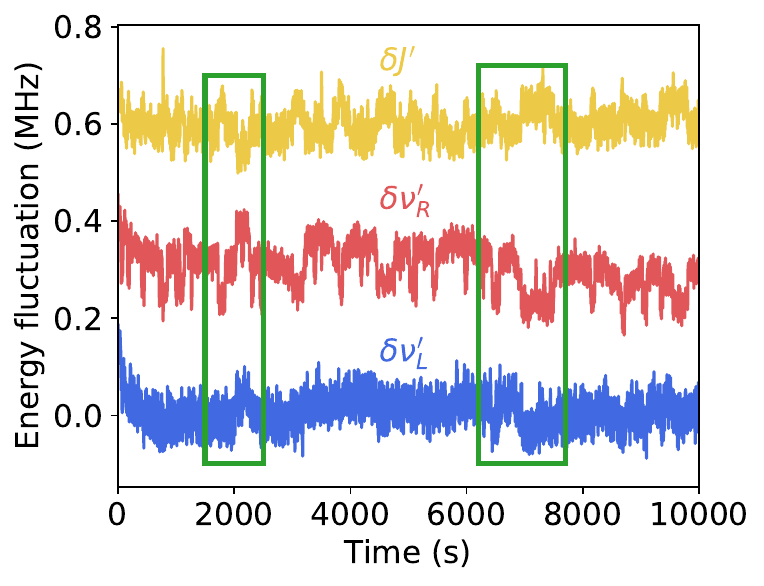}
\caption{Time traces of the estimated qubit energies and exchange. The traces have been shifted vertically for better visibility but each trace is relative to the average value: $\braket{\nu_L}=15537.59$ MHz and $\braket{\nu_R}=15841.31$ MHz for the qubit energies, and $\braket{J}=18.85$ MHz  for exchange. The green rectangles highlight sample instances where correlated jumps in the signals occur.}
\label{fig:time_traces}
\end{figure}

An interleaved Ramsey cycle is composed of four sub-cycles that probe all $\nu_\alpha^\sigma$. The first (third) sub-cycle consists of initialization of the two qubits to the spin-down state, a $\pi/2$-rotation of the left (right) qubit, a free evolution time $t_f$, a second $\pi/2$-rotation of the left (right) qubit, and finalizes with readout of both spins. The second (fourth) sub-cycle consists of the same sequence but initializing the right (left) qubit to the spin-up state by applying a $\pi$-rotation. We repeat the interleaved Ramsey cycle with $t_f$ ranging between 0.04 $\mu$s to 4 $\mu$s in 0.04 $\mu$s steps to form a record from which we extract the four $\nu_\alpha^\sigma$ via Bayesian estimation \cite{Delbecq2016, Nakajima2020}. From those four energies we calculate estimates of the three quantities of interest: $J\equiv(\nu_L^\uparrow-\nu_L^\downarrow+\nu_R^\uparrow-\nu_R^\downarrow)/2$ and $\nu_\alpha\equiv(\nu_\alpha^\uparrow+\nu_\alpha^\downarrow)/2$. A sample of the obtained time traces is displayed in Fig.~\ref{fig:time_traces}, where the switching behavior characteristic of coupling to TLFs is observed as well as the presence of correlations. We also define the quantity $Z\equiv(\nu_L^\uparrow-\nu_L^\downarrow-\nu_R^\uparrow+\nu_R^\downarrow)/2$ that quantifies the errors in the estimation and is used to correct the noise spectra from the effect of these errors (see below and Ref.~\cite{Yoneda2023}). A sub-cycle takes 4.265 ms to be completed, so we are able to extract a single estimate of the set $\{\nu_L,\nu_R,J\}$ in 1.706 s. We measure $10^4$ records, meaning that we probe the noise for 4.739 hours in total.

\section{Correcting PSDs for the effect of estimation errors}\label{app:errors}

The Bayesian estimation of correlation functions used in this paper allows us to calculate PSDs and provides probability distributions indicating the degree of confidence. This confidence refers to the statistical accuracy at which we can estimate the PSD based on the data. Thus, it does not include any confidence of the input data itself. The data, in our case, are the $\nu_\alpha^\sigma$ estimated via Bayesian estimation of qubit energies \cite{Delbecq2016,Nakajima2020} from Ramsey measurements. Since the data comes from an estimation process, it has itself a degree of confidence that needs to be taken into account. One hopes an estimated quantity to equal the true value of the variable but, most likely, estimation and reality differ. We refer to the difference between an estimated variable and its true value as \textit{estimation error}. In this section we show how we correct the noise spectra for the effect of estimation errors. We follow the same procedure as in Refs.~\cite{Yoneda2023,Gutierrez-Rubio2022} and include it here for completeness.

\begin{figure}[htbp]
\centering
\includegraphics[width=0.45\columnwidth]{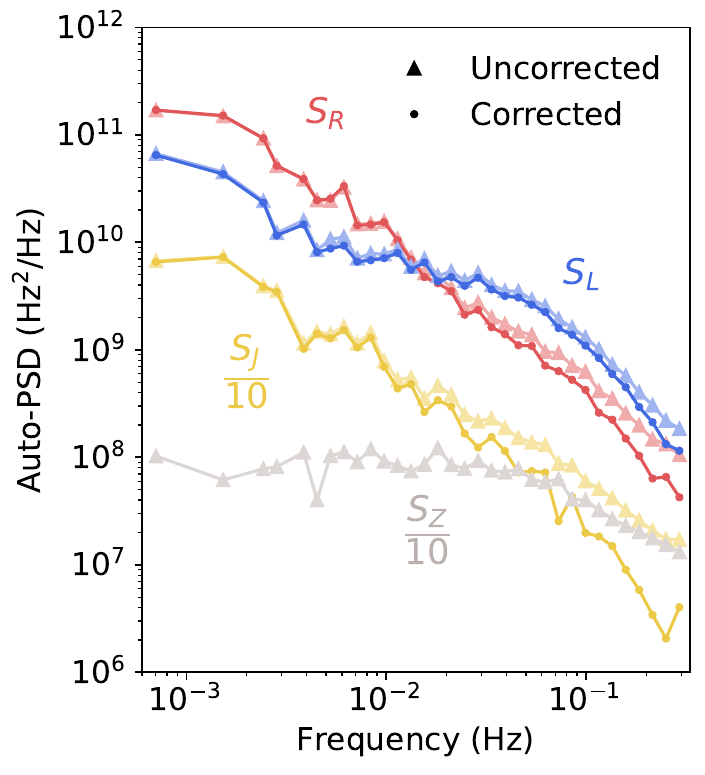}
\caption{Auto-PSDs with (colored dots) and without (faint colored triangles) the correction for estimation errors. We also include the auto-PSD of $Z$ to show how it affects the auto-PSD of $J$. Note that for $Z$ only the uncorrected version is plotted as $Z$ itself is our measure of the estimation errors, and its ``corrected'' version equals zero. We do not plot the probability distributions to not overcrowd the plot and we shift the auto-PSDs of $J$ and $Z$ downwards by a factor of 10 to avoid overlaps and improve visibility.}
\label{fig:corrected}
\end{figure}

Throughout this section we use apostrophes to distinguish between an estimated variable and its true value. In this way, $\nu_\alpha^{\prime\sigma}$ refers to our estimation of $\nu_\alpha^\sigma$. We define the estimation errors $\varepsilon_\alpha^\sigma$ by how much they differ: $\varepsilon_\alpha^\sigma\equiv\nu_\alpha^{\prime\sigma}-\nu_\alpha^\sigma$. To proceed further we assume that the errors are uncorrelated with each other, since each estimation is done separately, and with the true variables.

Given that we estimate four different $\nu_\alpha^{\prime\sigma}$, from which we extract only three quantities $\{\nu_L^\prime,\nu_R^\prime,J^\prime\}$, the system is overdetermined. We use the extra information to quantify the estimation errors via $Z^\prime\equiv(\nu_L^{\prime\uparrow}-\nu_L^{\prime\downarrow}-\nu_R^{\prime\uparrow}+\nu_R^{\prime\downarrow})/2$, which should equal zero were the estimations perfect. With the above, it is easy to prove that the auto-PSD of the estimated $J^\prime$ is:
\begin{align}
S_{J^\prime}(f)=S_J(f)+S_{Z^\prime}(f).
\end{align}
Hence, due to the presence of estimation errors, the auto-PSD $S_{J^\prime}$ we calculate from the estimated $J^\prime$ differs from the one of the true variable by $S_{Z^\prime}$. Therefore, we can obtain the PSD of interest as $S_J(f)=S_{J^\prime}(f)-S_{Z^\prime}(f)$, which is the result we plot in yellow in Fig.~3 in the main text. We note that given the Bayesian nature of our protocol to calculate PSDs, we deal with probability distributions instead of a collection of points. For this reason, correcting the spectra is not just a simple subtraction but rather a convolution of two probability distributions as outlined in Ref.~\cite{Gutierrez-Rubio2022}. Although more complicated, the benefit is that the corrected PSD also has associated a probability distribution from which we can infer a degree of confidence.

A similar procedure can be done to the auto-PSDs of the qubits and we get that the corrected auto-spectra are:
\begin{subequations}
\begin{align}
S_{L}(f)&=S_{L^\prime}(f)-\frac{1}{2}\left(S_{Z^\prime(f)}+C_{J^\prime Z^\prime}(f)\right),\\
S_{R}(f)&=S_{R^\prime}(f)-\frac{1}{2}\left(S_{Z^\prime(f)}-C_{J^\prime Z^\prime}(f)\right),
\end{align}%
\label{eq:corrected_SLR}%
\end{subequations}
where we again see how we can use $Z^\prime$ to correct the estimation errors. The auto-PSDs obtained from Eq.~\eqref{eq:corrected_SLR} are plotted in blue and red in Fig.~3. In order to highlight the importance of the correction of PSDs, in Fig.~\ref{fig:corrected} we plot the auto-PSDs both without and with the correction. There we can see that although at lower frequencies the spectra are mostly unchanged, it is at the higher frequencies where estimation errors overestimate the auto-PSDs. 

When correcting for the cross-PSDs, one gets that the unnormalized $C_{L^\prime R^\prime}(f)$ is unaffected by estimation errors: $C_{L^\prime R^\prime}(f)=C_{LR}(f)$. However, since the auto-PSDs are not immune to the errors, the normalized cross-PSD $c_{L^\prime R^\prime}(f)=C_{L^\prime R^\prime}/\sqrt{S_{L^\prime}(f)S_{R^\prime}(f)}$ needs to be corrected due to the presence of auto-PSDs in the denominator. In particular, since the auto-PSDs are overestimated, the normalized cross-PSD ends up being underestimated. For the correlations between qubit energies and exchange interaction we get:
\begin{subequations}
\begin{align}
C_{LJ}(f)&=C_{L^\prime J^\prime}(f)-C_{L^\prime Z^\prime}(f),\\
C_{RJ}(f)&=C_{R^\prime J^\prime}(f)+C_{R^\prime Z^\prime}(f),
\end{align}%
\end{subequations}
which shows that they are not immune to estimation errors, in contrast to $C_{L^\prime R^\prime}(f)$. However, their normalized version is also underestimated, similarly to $c_{L^\prime R^\prime}(f)$.

\begin{figure*}
\centering
\includegraphics[width=0.9\textwidth]{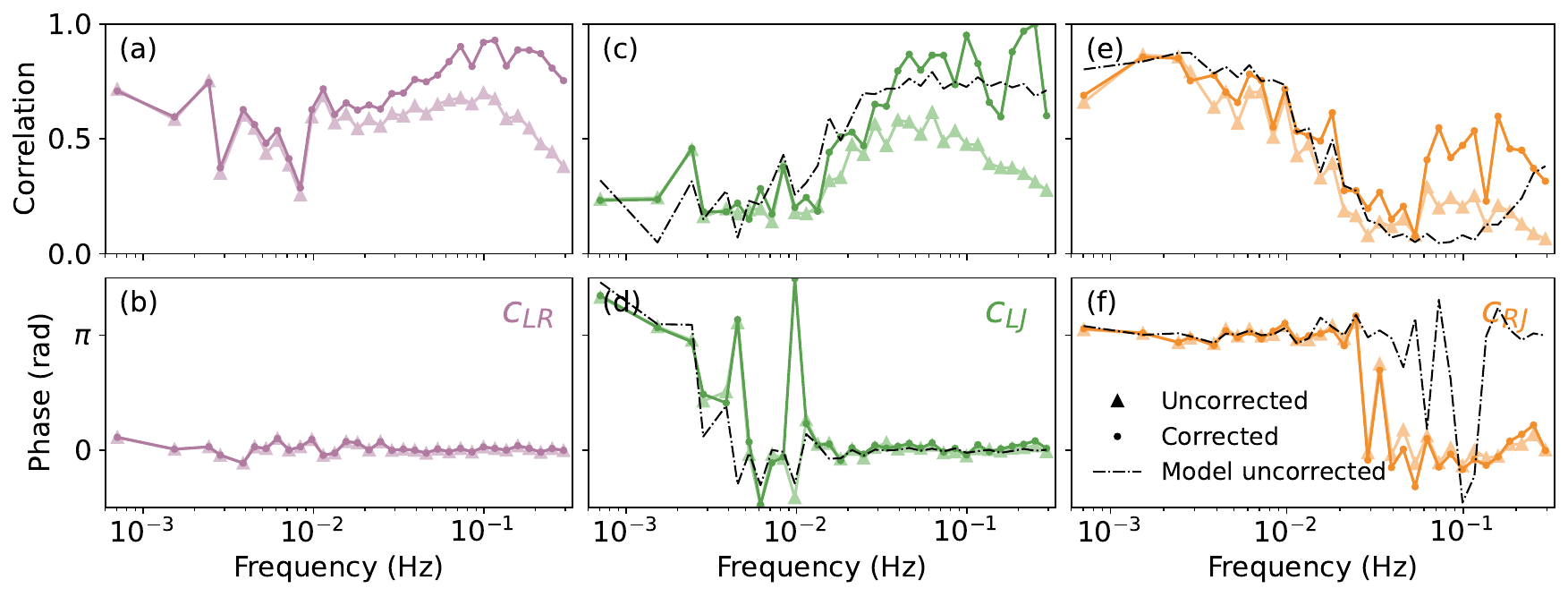}
\caption{Comparison between the corrected (colored dots) and uncorrected (faint colored triangles) normalized cross-PSDs. We omit the probability distributions to enhance visibility. The results from the model when using uncorrected PSDs as inputs is shown as black dashed lines, and shows disagreement with the data highlighting the need to correct the noise spectra.}
\label{fig:corrected_cross}
\end{figure*}

The correction of normalized cross-PSDs when dealing with the probability distributions of our Bayesian formalism is complicated and demands a lot of computational power \cite{Gutierrez-Rubio2022}. For this reason, when correcting normalized spectra we take a simple approach, where we drop the probability distributions and keep only the mean. Once done, dealing only with points, we calculate the corrected normalized cross-PSDs as: $c_{\alpha \beta}(f)=C_{\alpha \beta}/\sqrt{S_{\alpha}(f)S_{\beta}(f)}$. In Fig.~\ref{fig:corrected_cross} we plot both the corrected and uncorrected normalized cross-PSDs to show how crucial it is to correct the spectra. There, we see how the cross-PSDs can be severely underestimated at the higher frequencies. Furthermore, in that same figure we plot as black dashed lines the predictions of the model from Eq.~(3) of the main text, when using uncorrected spectra as inputs. The results from the model should be compared, then, with the uncorrected spectra plotted in triangles. We can observe large disagreements between model and data at higher frequencies, not only in magnitude but also in the phase. This emphasizes the need to correct the spectra to observe the excellent agreement of the charge noise model, as we did in Fig.~4. Without correcting the PSDs, the disagreements with the model would mistakenly suggest the presence of other noise mechanisms.

During the correction of normalized cross-PSDs, we dropped the probability distributions that gave us a confidence degree on the estimation. To still convey a degree of confidence we take advantage of the fact that the correction is of most importance at the higher frequencies. In Fig.~4 we plot the uncorrected normalized cross-PSDs including the probability distributions up to a certain threshold frequency. At frequencies higher than the threshold we plot the corrected spectra, which lacks the degree of confidence. Luckily for us, it is also at the higher frequencies that the PSDs have narrower probability distributions, see Figs.~3 and 4b in the main text. So, discarding the probability distributions and treating the PSDs as points (as done during the correction procedure) is justified.

\section{$J$ as a function of $\Delta$}\label{app:j_vs_d}

\begin{figure}[htbp]
\centering
\includegraphics[width=0.4\columnwidth]{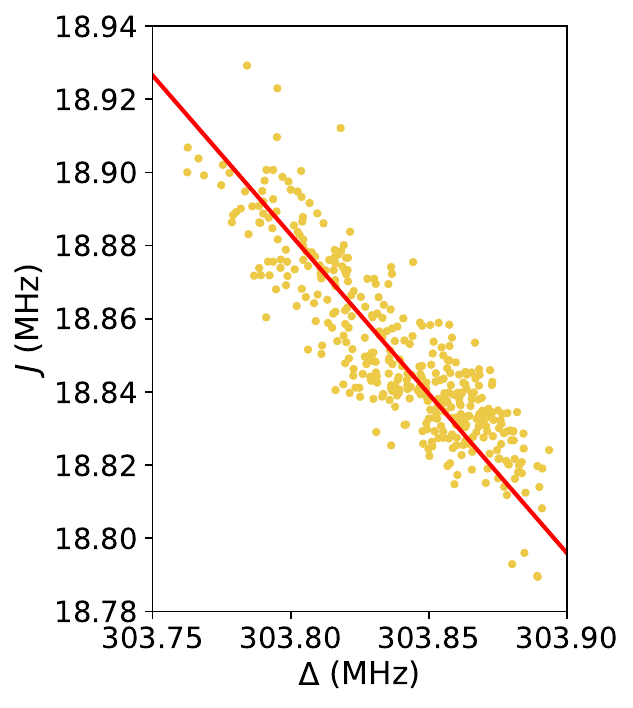}
\caption{$J$ as a function of $\Delta$. The red line shows a linear fit from which we extract $\partial J/\partial\Delta=-0.87$.}
\label{fig:j_vs_d}
\end{figure}

To evaluate the model of charge noise one needs the value of the derivative $\partial J/\partial\Delta$. We extract it from the data itself by plotting the estimated values of $J$ and $\Delta$ and performing a linear fit, as shown in Fig.~\ref{fig:j_vs_d}. From this we obtain $\partial J/\partial\Delta=-0.87$ allowing us to evaluate the model with all information coming from the data itself, without any free parameters.

\section{Two-TLF model of the PSDs}

In the main text, we identified two dominant TLFs located on the left and right sides of the qubit pair. While Fig.~3 shows the auto-PSDs of the qubit energies expected from two TLFs, the same procedure is in fact applied simultaneously to all PSDs. Here, we describe the fitting method in detail and present the full set of results.

\begin{figure}[htbp]
\centering
\includegraphics[width=0.45\columnwidth]{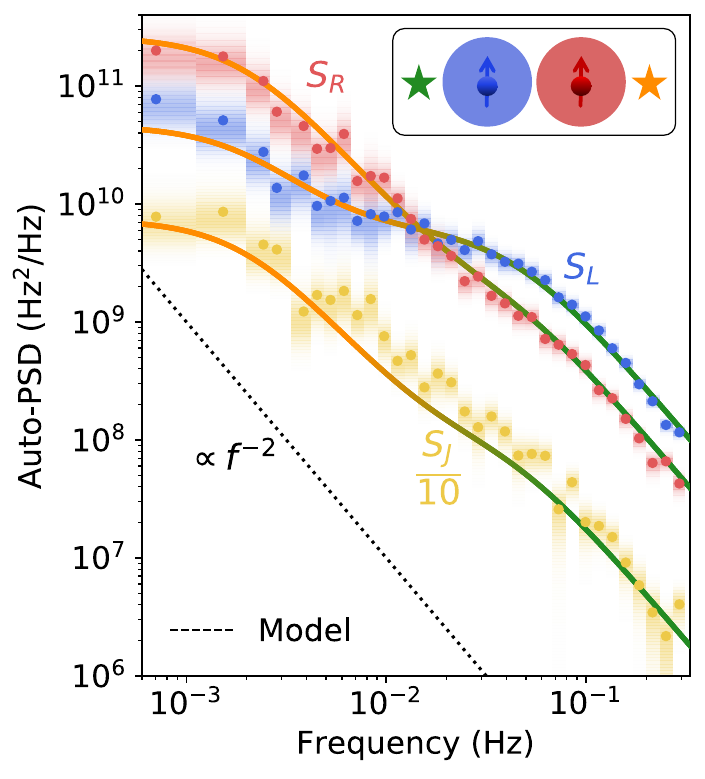}
\caption{Same as Fig.~3, but now also showing as curves the results of the two-TLF model for $S_J(f)$. The color gradient in the curves represent the contribution from each TLF, depicted as stars in the inset.}
\label{fig:auto-psds_fit}
\end{figure}

The energy fluctuation of qubit $\alpha$ can be modeled as the additive contribution from all TLFs: $\delta\nu_\alpha(t)=\sum_i g_i^\alpha q_i(t)$, where $g_i^\alpha$ is the coupling strength between TLF $i$ and qubit $\alpha$, and $q_i(t)$ switches between 1 or -1 denoting the state of TLF $i$, assumed to follow $\braket{q_i(t)q_j(t+\Delta t)}=\delta_{ij} \exp(-|\Delta t|/\tau_i)$, with $\tau_i$ the characteristic switching time. Defining the Lorentzian spectrum as $L(\tau_i,f)\equiv 2\tau_i/(1+4\pi^2 f^2 \tau_i^2)$, the auto-PSD of qubit $\alpha$ becomes $S_\alpha(f)=\sum_i (g_i^{\alpha})^2 L(\tau_i,f)$, while the cross-PSD between the two qubits is given by $C_{LR}(f)=\sum_i g_i^L g_i^R L(\tau_i,f)$.

\begin{figure}
\centering
\includegraphics[width=0.9\columnwidth]{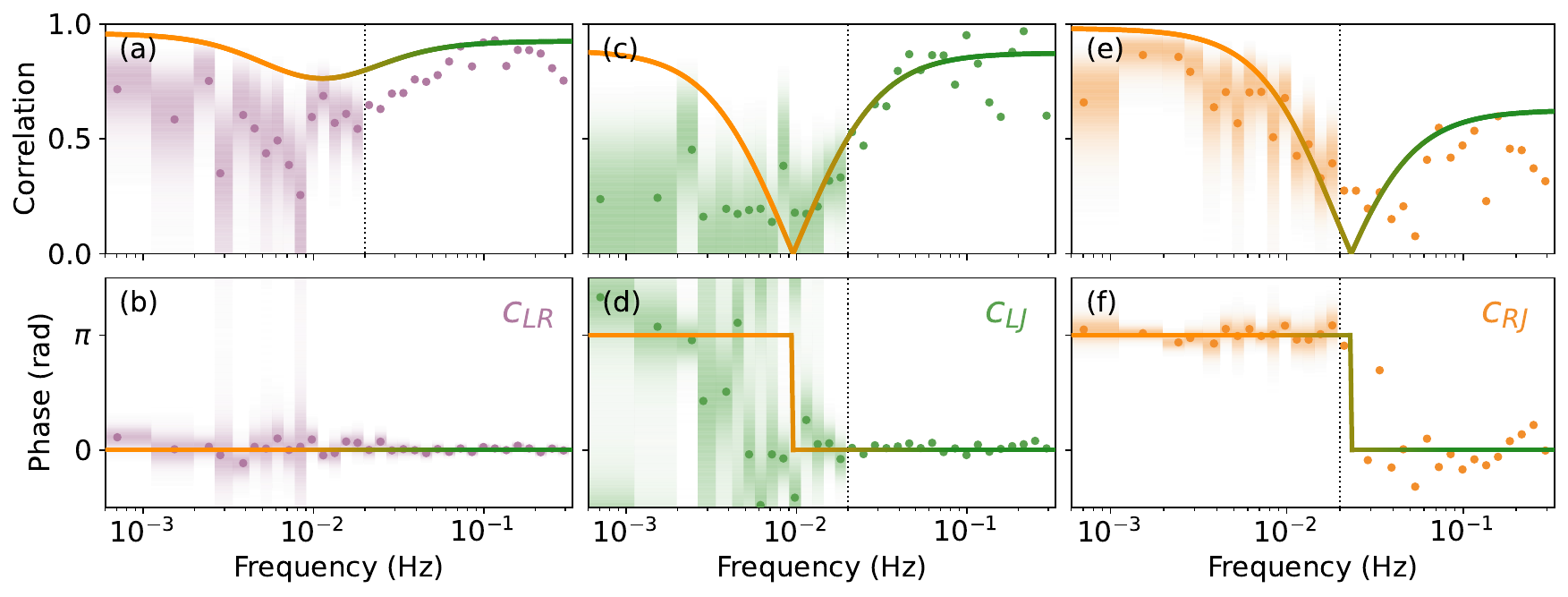}
\caption{Same as Fig.~4, including the results of the model with two-TLFs as curves. The color gradient in the curves signal the contribution from each TLF, schematically depicted as stars in the inset of Fig.~\ref{fig:auto-psds_fit}.}
\label{fig:cross-psds_fit}
\end{figure}

We performed a simultaneous fit of $S_L$, $S_R$, and $C_{LR}$, using as free parameters the coupling strengths and switching times for the two TLFs. The remaining spectra, $S_J$, $C_{LJ}$, and $C_{RJ}$, were then obtained using Eq.~(3) of the main text. From the fit we extract $|g_1^L|=28.5$ kHz, $|g_1^R|=15.6$ kHz, and $\tau_1=3.7$ s for one TLF, and $|g_2^L|=14.9$ kHz, $|g_2^R|=38.0$ kHz, and $\tau_2=90$ s for the other one. The results of the TLF model are shown in Figs.~\ref{fig:auto-psds_fit} and \ref{fig:cross-psds_fit}, where the gradient-colored curves represent the contributions from the two TLFs. The color encodes which TLF dominates at a given frequency: orange corresponds to the TLF on the right, and green to the one on the left, as indicated in the inset of Fig.~\ref{fig:auto-psds_fit}. 
In general, the model shows good agreement with the measurements, confirming that the main features in the PSDs are captured by two dominant TLFs. Small deviations from the model suggest the presence of additional TLFs contributing to the noise; however, these are weaker and do not give rise to strong correlations, likely because they are situated farther away from the qubits.

\section{Cross-PSDs from Ref.~\cite{Yoneda2023}}

\begin{figure}[htbp]
\centering
\includegraphics[width=0.9\columnwidth]{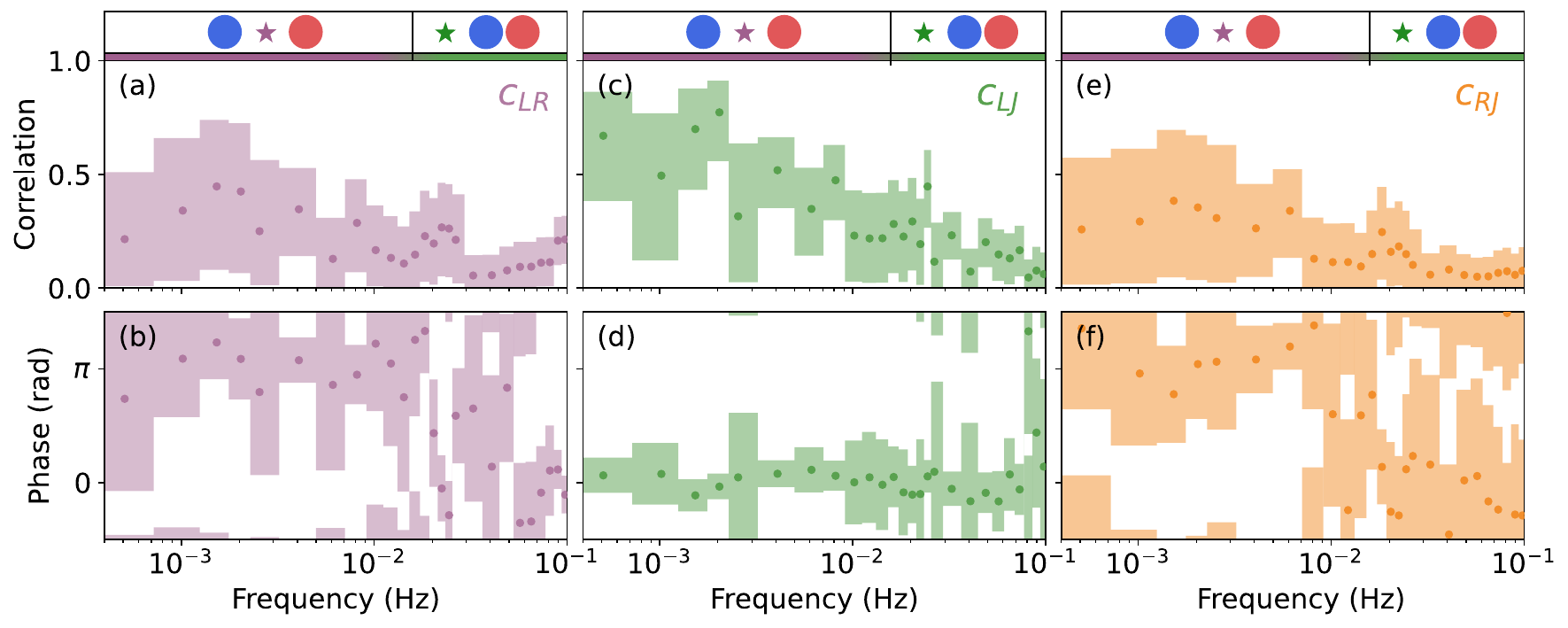}
\caption{Normalized cross-PSDs for the two-qubit system of Ref.~\cite{Yoneda2023}. $c_{LR}$ is shown in purple in (a) and (b), $c_{LJ}$ in green in (c) and (d), and $c_{RJ}$ in orange in (e) and (f). The normalized magnitudes are shown in (a), (c), and (e), while the corresponding phases are shown in (b), (d), and (f). In all panels, the shaded regions depict 90\% confidence intervals extracted from the probability distribution of the Bayesian estimated cross-PSD with the mean marked as points. The schematics on top indicate the regimes of dominant TLF: a TLF in between the qubit pair at low frequencies, and a TLF to the left at high frequencies.}
\label{fig:yoneda2023}
\end{figure}

In this section, we apply our noise-source triangulation method to the experimental data from Ref.~\cite{Yoneda2023}. In Fig.~\ref{fig:yoneda2023}, we reproduce the cross-PSD data shown in Figs.~2 and 4 of that work. Two distinct frequency regimes can be identified. At frequencies above 15 mHz, all cross-PSDs are positive (with phase equal to zero), consistent with the configuration shown in Fig.~2a of the main text, where the TLF is located to the left of the qubits. In contrast, at frequencies below 15 mHz, the sign of the cross-PSD $C_{LR}(f)$ becomes negative (with phase equal to $\pi$), indicating a TLF situated between the qubits, as depicted in Fig.~2c of the main text. The cross-PSDs involving exchange interaction also support this interpretation, with $C_{LJ}>0$ and $C_{RJ}<0$. This case, which did not appear in the device analyzed in the main text, completes the set of scenarios outlined in Fig.~2.

\section{Model of TLF-qubit coupling for triangulation}

The microscopic nature of TLFs in solid-state devices remains an open question, with possible origins including charge traps, dangling bonds, hydrogen impurities, and tunneling atoms \cite{Muller2019,Lisenfeld2016}. Each candidate implies a different functional form for the qubit-TLF interaction. As an illustrative example, we consider elementary charge trapping/detrapping events as the TLF transitions, highlighting how triangulation can be made more precise when a physical model is available. Importantly, PSD measurements across multiple qubits refine such a model, since it must reproduce all observed correlations simultaneously, thereby providing a pathway to identify the underlying TLF mechanism.

The qubits in the device of Fig.~1 of the main text are electron spins subject to an engineered magnetic gradient, designed to satisfy $\partial B_z/\partial x \approx 0.1$ mT/nm and $\partial B_z/\partial z = 0$ at the qubit positions \cite{Rojas-Arias2023}. For simplicity, we assume parabolic confinement with orbital energy $\hbar\omega_0=1.5$ meV, consistent with device modeling and previous measurements \cite{Camenzind2019}. A fluctuation of the electric field $\delta\vec{E}$ at the position of qubit $\alpha$ due to a TLF state switch displaces the electron wavefunction and produces a Zeeman energy shift according to Eq.~(1) of the main text. Explicitly,
\begin{align}
\delta\nu_\alpha = \frac{g\mu_B}{h}\frac{\partial B_z}{\partial x}\frac{e}{m\omega_0^2}\delta E_x,
\end{align}
where $e$ is the electron charge, $m=0.19m_0$ the effective electron mass in silicon, and $m_0$ the free-electron mass.

We assume the TLFs reside in the middle of the oxide layer above the quantum well, specifically the 2 nm SiO$_2$ below the metallic gates. Screening by the gates makes the TLF effectively a charge dipole $p$ oriented perpendicular to the 2DEG, located at height $y_0=52$ nm (50 nm SiGe spacer plus 2 nm oxide). The in-plane coordinates of the TLF are denoted $(x,z)$. The corresponding electric-field component at qubit $\alpha$ is
\begin{align}
\delta E_x = \frac{3\ p\ y_0\ x}{4\pi\epsilon\epsilon_0\big[(x-x_\alpha)^2 + y_0^2 + (z-z_\alpha)^2\big]^{5/2}},
\label{eq:dipole}
\end{align}
with $\epsilon$ the effective dielectric constant of the heterostructure (taken as $13$), $\epsilon_0$ the vacuum permittivity, and $p = e l_y$ the dipole moment, where $l_y=2$ nm corresponds to the oxide thickness.

Using the dipole model  from Eq.~\eqref{eq:dipole}, we generate the contour plots in Fig.~5b and identify the TLF positions where the qubit energy shifts $|\delta\nu|$ are simultaneously consistent with the values extracted from PSD fits.

\section{Construction of tiling plots for TLF localization}

In the main text we introduced the idea of ``tiling'' real space into distinct regions where a fluctuator may reside, consistent with the signs of cross-PSDs extracted from qubit measurements. This construction is conceptually simple: for each hypothetical TLF position $\vec{r}$, one computes the qubit response to a telegraph switch, assigns its sign, and then evaluates the set of cross-PSD signs between qubit pairs. Points in space that lead to the same signature are grouped into a tile. In this section we present the models underlying this procedure and discuss how the inclusion of different physical ingredients (non-uniform magnetic gradients, exchange coupling, and anisotropic $g$-tensors) enriches the tiling.

\begin{figure}[htbp]
\centering
\subfloat{\includegraphics[width=0.8\columnwidth]{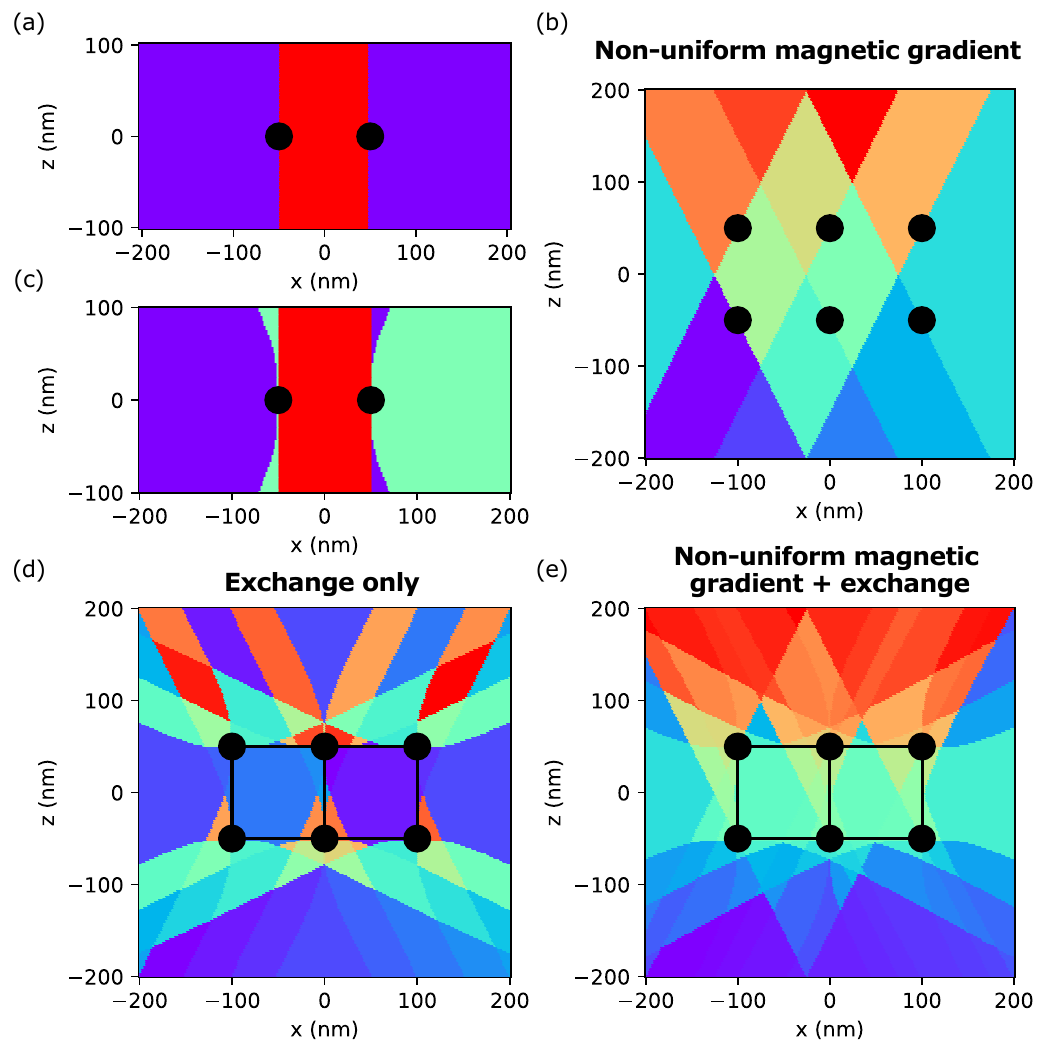}\label{fig:tiles_2x1}}
\subfloat{\label{fig:tiles_3x2}}
\subfloat{\label{fig:tiles_2x1_exch}}
\subfloat{\label{fig:tiles_3x2_exch}}
\subfloat{\label{fig:tiles_3x2_all}}
\caption{Different types of space tilings based on cross-PSD signs. (a) The simple case of two qubits in a uniform magnetic gradient in the $x$ direction. Taking only qubit energies into account, the space is divided into two regions which can be distinguished based on the signs of the cross-PSDs. (b) Generalization of the tiling in (a) for the case of a $3\times2$ qubit array in the presence of a non-uniform magnetic gradient $\nabla B_z\propto \hat{x}-0.1z\hat{z}$, forming a tangram-like pattern. (c) Same as (a) but including exchange into the set of variables. With it, it becomes possible to distinguish whether the TLF is to the left or right of the qubits, following the classification from Fig.~2 in the main text. (d) Tiling for the $3\times2$ qubit array when only correlations of exchange interactions are taken into account, corresponding to the situation relevant for EO qubits. Black lines indicate the exchange couplings considered during the classification. (e) General tiling of the $3\times2$ qubit array when including both a non-uniform magnetic gradient and exchange couplings. Closer inspection reveals that the pattern combines features from those in (b) and (d). In all plots the black circles indicate qubits and the different colors signal regions with unique cross-PSD signatures.}
\label{fig:tiles_general}
\end{figure}

\subsection{Space tiling from cross-PSDs of qubit energies}

We begin with the case considered in the main text of electron spins subject to a magnetic-field gradient. A fluctuation in the electrostatic environment displaces the qubit wavefunction and modifies its Zeeman splitting. To leading order this shift is
\begin{align}
\delta\nu_\alpha = \frac{g\mu_B}{h}\,\nabla B_z(\vec{r}_\alpha)\cdot \delta \vec{r}_\alpha,
\label{eq:delta_nu_basic}
\end{align}
where $\nabla B_z(\vec{r}_\alpha)$ is the local field gradient and $\delta\vec{r}_\alpha$ is the displacement of the wavefunction center. For tiling purposes only the sign $s_\alpha$ of $\delta\nu_\alpha$ is required. Since the displacement points towards or away from the TLF, we write
\begin{align}
s_\alpha(\vec{r}) = \mathrm{sign}\!\Big[\nabla B_z(\vec{r}_\alpha)\cdot (\vec{r}-\vec{r}_\alpha)\Big],
\label{eq:signrule}
\end{align}
with $(\vec{r}-\vec{r}_\alpha)$ the vector from the qubit to the TLF. The cross-PSD sign between qubits $\alpha$ and $\beta$ is then
\begin{align}
\mathrm{sign}\left[C_{\alpha\beta}(\vec{r})\right] = s_\alpha(\vec{r})\,s_\beta(\vec{r}).
\label{eq:crosspsd}
\end{align}

Equations~\eqref{eq:signrule} and \eqref{eq:crosspsd} partition the plane into regions of constant signature. For two qubits under a uniform gradient this leads to only two distinct possibilities: the TLF is located between the qubits, or to either side. This elementary partition is illustrated in Fig.~\ref{fig:tiles_2x1}. Although simple, this construction already shows how cross-PSD information restricts possible fluctuator positions.

In realistic devices the magnetic field cannot always be taken as a uniform gradient. Nevertheless, the rule in Eq.~\eqref{eq:signrule} is general: one evaluates the gradient $\nabla B_z(\vec{r}_\alpha)$ from simulations or measurements of the micromagnet profile, and uses it to determine $s_\alpha(\vec{r})$. Curvature and even sign changes in $\nabla B_z$ enrich the tile boundaries, producing nontrivial patterns such as those in Fig.~\ref{fig:tiles_3x2}, for the case of a larger $3\times2$ array. Thus, the tilings encode both the fluctuator geometry and the device-specific magnetic field landscape.

\subsection{Including exchange correlations}

The description slightly changes when exchange coupling is considered. A TLF-induced spatial shift displaces the qubits asymmetrically. One qubit is pushed more strongly than the other, changing their relative separation $d$. To first order the exchange fluctuation is
\begin{align}
\delta J(\vec{r}) = \frac{\partial J}{\partial d}\,\delta d(\vec{r}), 
\label{eq:deltaJ}
\end{align}
with $\partial J/\partial d<0$, since the qubits' wavefunction overlap decreases when they grow further apart. The corresponding sign is
\begin{align}
s_J(\vec{r})=\mathrm{sign}[\delta J(\vec{r})].
\end{align}
Together with $s_\alpha(\vec{r})$, this generates a richer set of correlators. For the case of two qubits the set is
\begin{align}
\{s_L s_R,\, s_L s_J,\, s_R s_J\},
\end{align}
which in practice resolves whether the TLF lies to the left or to the right of the qubit pair. The tiling is therefore refined, as we show in Fig.~\ref{fig:tiles_2x1_exch}. With this calculation, in which we model the TLF-qubit repulsion to be $\propto |\vec{r}-\vec{r}_\alpha|^3$, we confirm our classification from Fig.~2 of the main text. This illustrates how including additional couplings systematically increases the discriminating power of the method. Additionally, we observe the presence of two narrow bands near the qubits to each side, in which the cross-PSD signatures are compatible with those of the major area on the opposing side. In these bands, the repulsion on one of the qubits acts mostly vertically, rather than horizontally, causing the interqubit separation to increase despite the TLF being on one side of the qubit pair. This does not nullify our classification but rather enriches it, noting that any ambiguity in TLF position is easily clarified by including information from other qubits in the array.

\subsection{Exchange-only qubits}

For exchange-only (EO) qubits there are no magnetic field gradients, logical states are insensitive to individual qubit displacements and only fluctuations of the exchange couplings matter. The relevant quantities are therefore
\begin{align}
s_{J_i}(\vec{r}) = \mathrm{sign}\!\big[\delta J_i(\vec{r})\big], \qquad 
\mathrm{sign}\left[C_{J_iJ_j}(\vec{r})\right] = s_{J_i}(\vec{r})\,s_{J_j}(\vec{r}),
\end{align}
where $J_i$ label the different exchange couplings in the array. The set of all $\{s_{J_i}s_{J_j}\}$ constrains the plane and yields a tangram-like partition, as we show in Fig.~\ref{fig:tiles_3x2_exch} for a $3\times2$ array, where we are only considering exchange couplings between nearest-neighbor spins (black lines). Naturally, when considering only exchange interaction, the tiling of space follows a different geometry than when considering only qubit energy fluctuations (Fig.~\ref{fig:tiles_3x2}). When characterization as in the main text is available, with simultaneous access to qubit energies and exchange, one can construct the tiling of Fig.~\ref{fig:tiles_3x2_all} using the full set of variables. Although complicated at first sight, upon closer inspection it is clear that it is a combination of the tilings from Figs.~\ref{fig:tiles_3x2} and \ref{fig:tiles_3x2_exch}.

\subsection{Hole spins with anisotropic $g$-tensor}

For hole qubits the coupling mechanism differs qualitatively from that of electrons. The Zeeman splitting is governed by the full $g$-tensor,
\begin{align}
E_Z = \frac{\mu_B}{2}\,\vec{B}\cdot \mathbf{g}_\alpha\cdot\vec{\sigma}_\alpha ,
\end{align}
where $\mathbf{g}_\alpha(\vec{r}_\alpha)$ depends on the local confinement environment. We can always define an effective magnetic field $\vec{B}_\mathrm{eff}(\vec{r}_\alpha)\equiv\vec{B}\cdot\mathbf{g}_\alpha$, such that a TLF-induced qubit displacement $\delta\vec{r}_\alpha$ yields the change in qubit energy
\begin{align}
\delta\nu_\alpha=\frac{\mu_B}{h}\nabla B_\mathrm{eff}^{(z)}\cdot\delta\vec{r}_\alpha,
\end{align}
where we have defined $z$ as the quantization axis for the hole spin. We have reached the same expression as Eq.~\eqref{eq:delta_nu_basic}, showing the direct applicability of our method to holes, as long as the $g$-tensor of the qubits is characterized. We worked with fluctuations of the qubit position but changes in $g$-tensor from electric potential fluctuations can also be considered. Due to the natural randomness of typical hole-qubit devices we refrain from making a (perhaps meaningless) plot, but note that in a well-characterized system \cite{John2025} the tiling of space for TLF location remains possible.

In summary, the tiling construction rests on a universal principle: assign to each TLF location the signs of the induced qubit or exchange fluctuations, form their cross-products, and identify regions of constant signature. Starting from the simple case of two qubits in a uniform magnetic-field gradient, we have shown how additional ingredients (exchange correlations, arbitrary magnetic fields, and anisotropic $g$-tensors) progressively enrich the tilings and thereby enhance the ability to triangulate the noise sources.


\end{document}